\documentclass[preprint,12pt]{elsarticle}

\usepackage{graphics} 
\usepackage{graphicx} 
\usepackage{amssymb} 
\usepackage{amsthm}  
\usepackage{lineno}

\journal{Nuclear Instruments and Methods in Physics Research Section A }

\begin{document}

\begin{frontmatter}



\title{Development of a plasma panel radiation detector}


\author[label1]{ R.~Ball }\author[label2]{J.~R.~Beene }
\author[label3]{M.~Ben-Moshe}  \author[label3]{Y.~Benhammou} \author[label3]{R.~Bensimon} 
\author[label1]{ J.~W.~Chapman} 
\author[label3]{E.~Etzion} 
\author[label1]{ C.~Ferretti} 
\author[label4]{P.~S.~Friedman}
\author[label1]{ D.~S.~Levin} 
\author[label3]{{Y.~Silver}} 
\author[label2]{ R.~L.~Varner}
\author[label1]{C.~Weaverdyck}
 \author[label1]{ R.~Wetzel} 
\author[label1] { B.~Zhou}
\author[label5] { T.~Anderson}
\author[label5] { K.~McKinny}
\author[label6]{E. H. Bentefour}

\address[label1]{ University of Michigan, Department of Physics, Ann Arbor, Michigan, 48109, USA.}
\address[label2]{ Oak Ridge National Laboratory, Physics Division, Oak Ridge, Tennessee, 737831, USA.}
\address[label3]{ Tel Aviv University, Beverly and Raymond Sackler School of Physics and Astronomy, Tel Aviv, Israel.}
\address[label4]{ Integrated Sensors, LLC, Ottawa Hills, Ohio, 43606, USA.}
\address[label5]{GE Measurement and Control, Reuter-Stokes Product Line, Twinsburg, Ohio, 44087, USA.}
\address[label6]{ Ion Beam Applications S. A., Louvain La Neuve, B-1348 Belgium.}

\begin{abstract}
 
This article reports on the development and experimental results 
of commercial plasma display panels adapted for their potential 
use as micropattern gas radiation detectors.
The plasma panel sensor (PPS) design and materials include glass substrates, 
metal electrodes and inert gas mixtures which provide a physically robust, 
hermetically-sealed device.
Plasma display panels used as detectors were tested with cosmic ray muons,
beta rays and gamma rays, protons, and thermal neutrons.
The results demonstrated rise times and time resolution of a few nanoseconds,
as well as sub-millimeter spatial resolution compatible with the pixel pitch.

\end{abstract}


\end{frontmatter}


\newpage

\section{Introduction}

This article reports on the development and testing of 
commercial plasma display panels adapted for 
use as a novel type of micropattern gas radiation detector.
Plasma display panels (PDP) are commonly used in televisions and graphical 
display devices. Their design and production is supported by an extensive and 
experienced industrial base with four decades of development. 
The application of PDPs as particle detectors is referred to as a 
plasma panel sensor (PPS)
\cite{PPS_NIM,OURPPSPAPER,PPSPAPER,OURPPSPAPER1,OURPPSPAPER2,IEEE2010}.
The primary motivation underlying the PPS concept is to use well 
established manufacturing processes of PDPs to develop scalable, inexpensive 
and hermetically sealed gaseous detectors with the potential for a broad 
range of commercial and research applications.

A commercial PDP consists of millions of cells per square meter, each of which 
can initiate and sustain a plasma discharge when addressed by a bias-voltage 
signal~\cite{PDPs}.  A PDP, in the simplest matrix configuration, consists of 
an envelope 
of two flat, parallel, glass plates with line electrodes deposited on 
the internal surfaces.  The plates are sealed together at the edges with the 
top and bottom electrodes aligned perpendicularly. The gap separating the two 
plates is between 200-400~$\mu$m and is filled with a Penning gas mixture of 
mostly Xe, Ar or Ne, typically at pressures of about one-half atmosphere.
In such a structure, a pixel is made of an electrode intersection and gas gap.

\begin{figure}[!ht]\centering
  \includegraphics[width=0.8\textwidth]{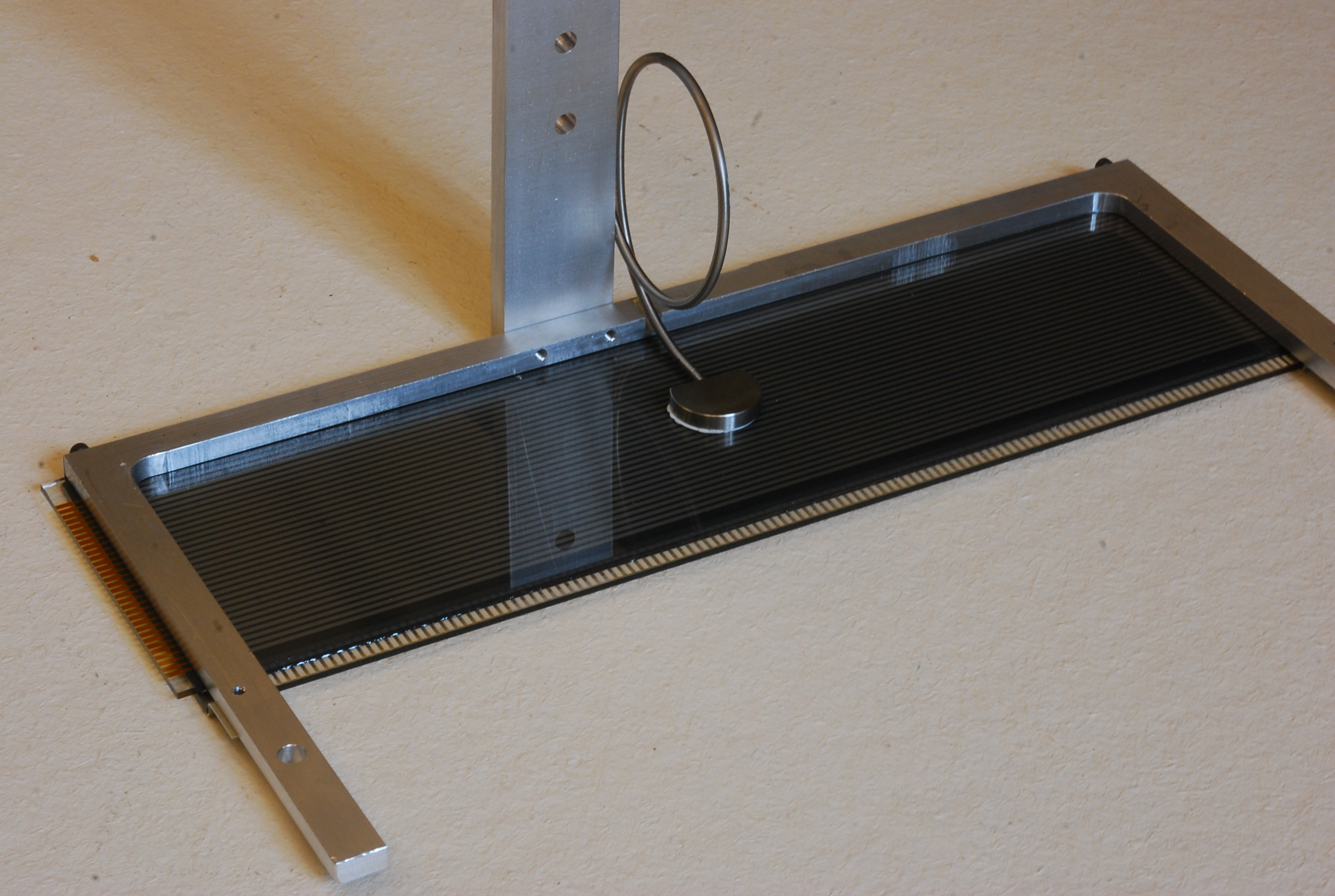}
  \vspace{-3mm}
  \caption{A modified commercial PDP 
           (type VPA in Table~\ref{table:PANEL_SPEC}). }
  \label{fig:pdp_pic}
\end{figure}

A PPS incorporates the general structure of a PDP, but instead of actively 
inducing a plasma discharge with an applied  voltage delivered to an addressed 
pixel, the plasma discharge is caused by ionizing radiation entering a PPS 
cell biased with  a constant DC voltage above the Paschen potential.
Results reported in this article were produced with commercial PDPs modified 
in specific ways to allow them to be used as particle detectors. 
They operated with DC bias voltages, had no dielectric barriers to isolate 
individual cells and had no phosphors in the cells. 
These panels were a simple matrix of anodes and cathodes
with a gas filled gap of a few hundred micrometers. An example of a test 
device is shown in Figure~\ref{fig:pdp_pic}.
The modifications made to the panels were:
\begin{itemize}
\item Panels were normally fabricated with a hermetically sealed glass port.
      This was replaced with a stainless steel valve assembly that connects 
      the panel to a gas mixing and vacuum pump system. 
\item The original Ne based panel gas was replaced with a test gas.
\item The electrodes in the commercial panels were made from both Ni and 
      SnO$_2$:F. 
      A selection of the tested PPS panels used all Ni electrodes.
      The Ni was found to be much more resistant than the SnO$_2$:F 
      to sputtering degradation.
\item The electronics required for panel display operation was replaced with 
      signal extraction circuitry on the readout anode electrodes. 
\item The high voltage bias was routed through quench resistors connected to 
      each cathode electrode. 
\end{itemize}

The above modifications allowed these commercial units to serve as a useful
test bench for the PPS concept. The modified commercial panels served as 
prototypes for investigations and the results obtained from them have 
informed the next generation of PPS panels, the subject of a future paper.  
Commercial PDPs are sealed devices designed to work for $10^5$ hours.
One panel filled with a Xe-based gas mixture at 600~Torr and hermetically 
sealed in 2003 produced signals when operated as a PPS seven years later,
clearly demonstrating the long term stability of the materials and gas mixture.

The panels evaluated in this report differed from each other in the electrode 
material, pixel density and size, and gas gap, that is separation of the anodes
and cathodes.  
For the hit position studies, a type MP plasma panel was employed with 
a pixel pitch/granularity of $1.02$~mm.  
For the investigation of panel timing, efficiency and response to various 
particle types 
a larger  VPA panel was used with a pixel pitch/granularity of $2.54$~mm. 
The specifications of these two different types of panels are summarized in 
Table~\ref{table:PANEL_SPEC}. 

\begin{table}[!th]
\centering \vspace{-3mm}
\caption{\footnotesize{\textsl{PDP manufacturer's specifications.  
The packing fraction is the ratio between the active pixel area and total area. 
The electrode length includes only the section inside the gas volume.
HV=high voltage, RO=readout. The dielectric mesh defines the pixel 
perimeter but does not act as a barrier between pixels.  
The panel type name is an internal identifier without global significance.}}}
\small
\begin{tabular}{lcc}\hline\hline
	                   &  Panel type VPA  & Panel type  MP \\
	\hline
	HV electrodes material     & Ni       & Ni       \\
	HV electrodes width (mm)   & 1.40     & 0.44     \\
	HV electrodes length(mm)   & 81       & 65       \\
	RO electrodes material     & Ni       & SnO$_2$:F\\
	RO electrodes width (mm)   & 1.27     & 0.71     \\
	RO electrodes length(mm)   & 325      & 131      \\
	Electrodes pitch (mm)      & 2.54     & 1.02     \\
	Active pixel area (mm$^2$) & 1.50     & 0.22     \\ 
	Packing fraction           & 23.5\%   & 22.0\%   \\
	Gas gap (mm)		   & 0.38     & 0.29     \\
	Glass thickness (mm)       & 2.23     &  2.23    \\
	Electrodes thickness (mm)  & 0.02     & 0.02     \\
	Dielectric mesh thickness (mm)   & 0.02     & 0.02     \\
	VISHAY product number	   & PD-­‐128G032-­‐1 &	PD-­‐128G064-­‐1 \\
	\hline\hline
	\end{tabular}
	\label{table:PANEL_SPEC}
\end{table}
Signals were investigated from panels filled with different gas mixtures at 
various pressures, exposed to radiation from radioisotope sources, particle 
beams and cosmic rays. 
The various mixtures were either commercially obtained or produced in our 
gas mixing system. 

\section{Operational principles}

The use of PDPs  as particle detectors requires 
that a charged particle will generate enough ion-pairs to 
initiate an avalanche leading to a discharge.
This operational mode must be beyond the proportional region in the Geiger 
region~\cite{KNOLL} of gas ionization to achieve the desired fast, 
high gain response. 
This mode of operation of a cell can yield copious photons, photoelectrons 
and metastable atoms which could, due to the lack of physical barrier between 
the pixels, propagate and spread the discharge to other pixels. 

Various mechanisms mitigate discharge regeneration in a PPS.
Small amounts of gases such as CO$_2$ or CF$_4$ are added to the primary 
gas to absorb the photons through non-radiative vibrational and rotational 
excitations.
Penning type mixtures~\cite{Penning} may also be used wherein the dopant 
gas has a first ionization energy level lower than the host gas excited 
states, e.g., CO$_2$ or CF$_4$ in Ar.
The Penning transfer process allows for collisional de-excitation of the 
long-lived, metastable states. 
Finally, the discharge is externally quenched with a large resistor connected 
in series to each pixel or chain of pixels on the high voltage (HV) electrode. 
The pixels are then allowed to recover on a time scale determined by the $RC$ 
time constant, where C is the effective pixel capacitance, effectively 
dominated by the choice of the quench resistor. This time constant should be 
commensurate with the time required for positive ions to neutralize and gas 
metastable species to decay~\cite{LIFETIMES}.

The detailed mechanism for the discharge process is uncertain. A classic 
Townsend gas avalanche is limited by space charge buildup that negates the 
applied electric field, known as the Raether limit~\cite{STREAMER1, STREAMER2}.
The gas discharges that occurred in PPS pixels produced charge in 
excess of the Raether limit.  This is shown in the next section 
where the signal is described by a simple capacitive discharge model. 
This model is valid after the discharge has fully evolved between the pixel 
electrodes. 
We do not yet have the complete description of the progression from avalanche 
to full discharge that is required to accurately predict the gas dependent 
signal evolution.

\subsection{Signal model}

\begin{figure}[!ht] \centering
  \includegraphics[width=0.8\textwidth]{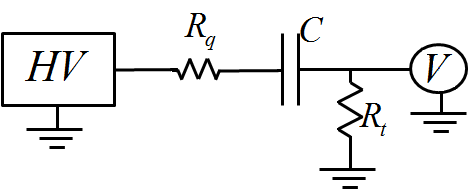}
  \vspace{-3mm}
  \caption{Idealized schematic view of a single pixel.}
  \label{fig:signal_extraction1}
\end{figure}

An idealized model for one pixel in the PPS detector is shown
in the equivalent circuit model in Figure~\ref{fig:signal_extraction1}, 
where $R_q$ is the quench resistor ($\sim 100 \, \rm{M}\Omega$), 
$C$ is the pixel total {\it effective} capacitance and 
$R_t$ is the $50 \, \Omega$ termination resistor over which the 
signal is read. 

\begin{figure}[!ht]\centering
  \includegraphics[width=0.8\textwidth]{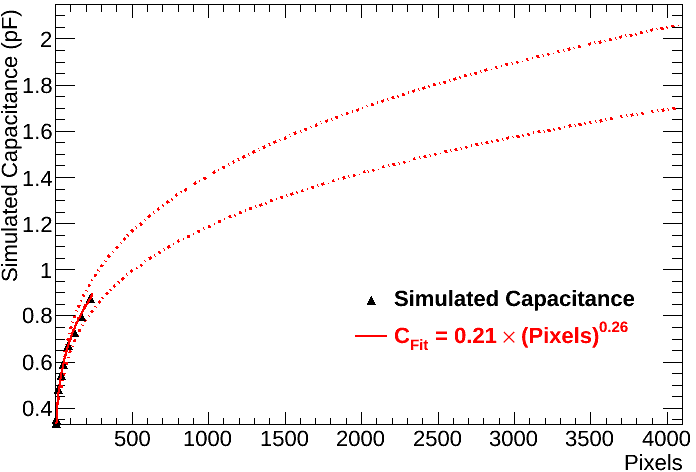}
  \vspace{-3mm}
  \caption{Computed capacitance versus the number of pixels.
           A fit (solid line) to the simulated data is extrapolated 
           with uncertainties (dashed lines) to the number of pixels
           in a type VPA panel.}
  \label{fig:SimCap}
\end{figure}

The effective pixel capacitance includes the capacitance presented by the 
crossing of two orthogonal electrode lines and that from stray or parasitic 
couplings with all the other electrodes in the panel.
This  effective pixel capacitance in a type VPA panel was modeled using the 
COMSOL~\cite{COMSOL} package which uses the finite element method to solve the 
Poisson equation. This three dimensional model simulates the entire volume of 
the detector and specifies the electrode 
dimensions, pitch, gap and number of electrodes in an orthogonal square array. 
The model did not account for the dielectric glass substrates, nor the 
thin dielectric mesh overlaid on the electrodes at the cell perimeter. 
The computation was run for a series of increasing array sizes up 
to $15 \times 15$ at which point further increases became computationally 
impractical. The simulated capacitance versus number of pixels, shown in
Figure~\ref{fig:SimCap}, was fit and extrapolated to the VPA panel array 
size of $32$ pixels $\times$ 128 pixels.  
The extrapolated result was $1.9\pm 0.15$~pF, where the uncertainty was 
determined  from the fit errors.  By comparison, the effective capacitance 
was also measured with an Agilent 4263B LCR meter. 
A series of ten measurements  were conducted for various pair combinations.
The measured value was $2.7 \pm 1.0 $~pF, consistent with the computation 
while the capacitance of a single electrode pair was calculated to be almost
an order of magnitude lower, at 0.35~pF.

An estimate of the amount of charge in a signal with an amplitude of several 
volts is given by: 
${  V_{signal} =\frac{\Delta q}{\Delta t} R_t}$ 
where  $V_{signal}$ is the measured signal amplitude, 
$\Delta t$ is  the pulse width and 
$R_t$ is the $50 \, \Omega$ termination resistance. 
Average values for MP type panels (see Table~\ref{table:PANEL_SPEC}) are 
$V_{signal}\approx 5$ volts and $\Delta t \sim$ 5 ns, yielding 
$\Delta q \sim 10^{-9} \, \rm{C}$, 
or about $10^{10}$ electrons.
Alternatively,   $\Delta q = C_{pixel} \cdot \Delta V_{HV}$ where 
$\Delta V_{HV}$ is the change in the bias voltage on one pixel during discharge.
 $\Delta V_{HV}$ is about 300 volts on top of a bias voltage of 1000 volts.
$C_{pixel}$ is the effective capacitance of the pixel described above, 
yielding a similar estimate of the charge. 
This amount of charge greatly exceeds $\approx 10^8$ gain expected from the 
Raether limit~\cite{STREAMER1, STREAMER2}.
\par
\begin{figure}[!ht]\centering
  \includegraphics[width=0.8\textwidth]{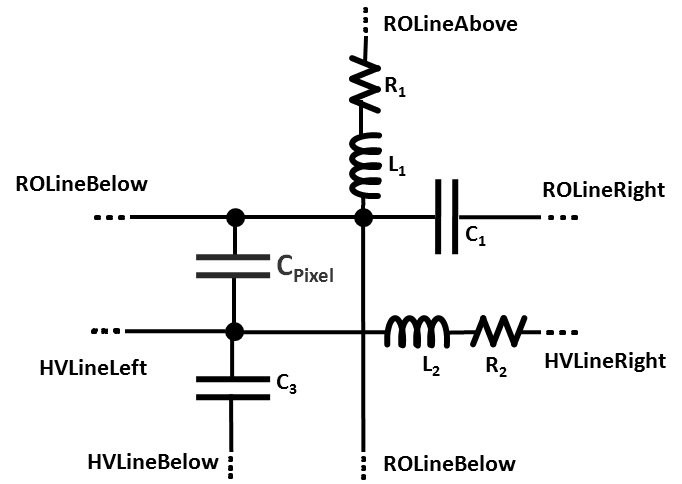}
  \vspace{-3mm}
  \caption{SPICE model of one cell of a commercial PDP.
            The parameters C, L and R are capacitances, inductances and
            resistances of a single cell coupled to it's neighbors.
            Dashed lines represent the cell's connections to other pixels. }
  \label{fig:SPICE_cell}
\end{figure}

The signal characteristics were further described by a SPICE~\cite{SPICE}  
model which simulated the electronic response of a single pixel embedded 
among neighbors, connected via direct and stray capacitances, inductances 
and line resistance. 
In this model the fundamental single pixel circuit included the capacitance 
of the cell, the self-inductance and resistivity of the lines, and the 
nearest neighbor parasitic coupling capacitances. 
Not included in the SPICE model were the readout electronics (printed 
circuit card, passive components, cables, etc).
The input parameters were determined with a COMSOL-based electrostatic model. 
Figure~\ref{fig:SPICE_cell} shows the schematic of the SPICE equivalent 
circuit model of one cell in the panel.
The physical development of the signal in the pixel was not modeled.
It was introduced as an impulse at capacitor $C_{pixel}$.

\begin{figure}[!ht]\centering
  \includegraphics[width=0.8\textwidth]{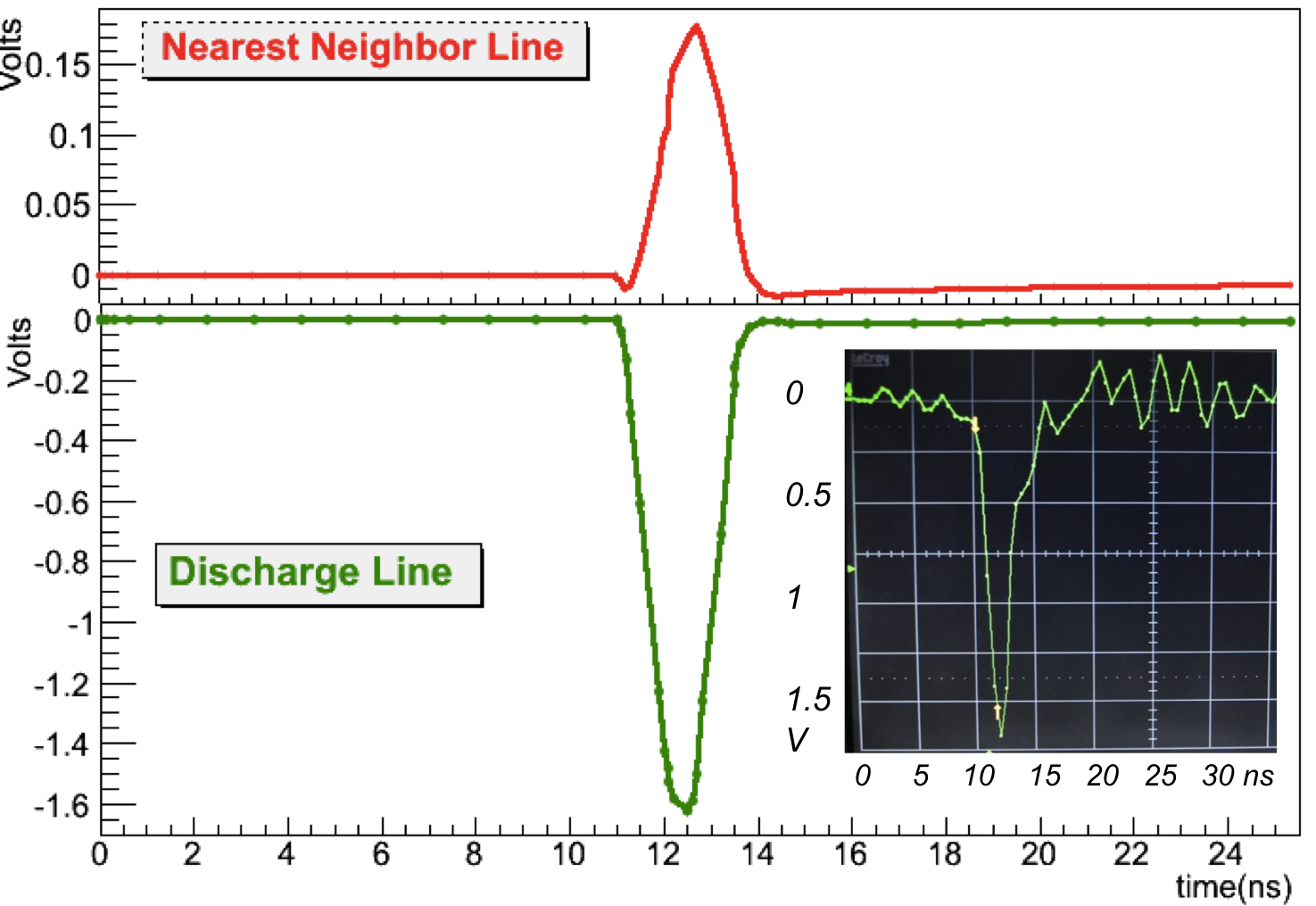}
  \vspace{-3mm}
  \caption{SPICE simulation results for the output pulse from a discharging
           cell (bottom) and the signal induced in an adjacent neighboring 
           electrode (top). The scope picture of a recorded signal in the
           bottom insert shows the same amplitude and time characteristic
           of the simulated signal.}
  \label{fig:SPICE_pulse}
\end{figure}

The full SPICE model connected all the neighboring cells to form a 
$ 5 \times 5 $ array of pixels.  Larger array sizes are computationally 
expensive and produce the same results.  Figure~\ref{fig:SPICE_pulse} 
shows the signal produced by the central pixel in the array and also 
a smaller transient pulse induced in an adjacent neighboring electrode.
The important qualitative attributes of the SPICE modeling included the pulse 
shape and the transient pulses on other lines. The signal induced in
neighboring pixels had greatly reduced amplitude and the opposite polarity 
of the discharging cell.

\subsection{Pixel dead time}
\label{DeadTime}
During the discharge, electric current starts to flow between the pixel 
electrodes.
This current results in voltage dropping on the quench resistor, reducing 
the electric field between the pixel's electrodes.  
In these conditions the pixel discharge is not sustained;
the discharge  terminates and the gas neutralizes.
The time required for the pixel electric field to regain a strength 
sufficient to enable a new discharge is determined by an RC time constant, 
where R is the quench resistance and C is the effective pixel capacitance.
For a quench resistance of 100 M$\Omega$ and approximately 2~pF effective 
capacitance, the RC time is $2\times10^{-4}$ seconds.  
The pixel dead time was experimentally determined to be about three times 
this RC time constant. 
For the modified commercial PDPs reported here, the quench resistance was
applied to an entire chain of 24 pixels serviced by one HV line, rather than 
a single isolated pixel. 

The RC time constants, and thus the pixel dead time, should be optimized 
according to the following considerations:
\begin{itemize}
\item If the dead time is not sufficiently long and the positive ion cloud 
      and/or the gas metastable states are not neutralized then the 
      discharge can restart without an external trigger, giving rise to
      more than one pulse per incident particle.
\item For gas recovery dead time longer than necessary, the observed 
      signal rate would saturate at an unnecessarily low rate and the panel 
      will lose efficiency at high irradiation rates.   
\end{itemize}
The optimal dead time changes with the gas content, with values in the 
range between 0.1~ms to a few milliseconds per pixel.

\section{Data Acquisition}

The characteristics of the pulses induced in the panel were measured with
a 1 gigasamples/sec (GSPS) digital oscilloscope or an evaluation board 
of the DRS4 chip~\cite{DRS4}. 
The evaluation board was equivalent to a four channel, 5 GSPS digital 
oscilloscope with control, readout and data storage performed by an 
external computer.

Extraction of the signals from the prototype PPS panels was done with a 
custom designed PCB readout card connected to the panel electrode pads.
The signal was picked off of a termination resistance ($R_{t}$ in 
Figure~\ref{fig:signal_extraction1}). 
These readout cards included attenuators for each readout line since 
the observed several volt signal amplitudes were too large for the readout 
electronics.

In order to instrument more readout lines and thus have a larger active
area in the panel, a more elaborate DAQ system was used. 
For this purpose a portable version of the readout system for the ATLAS 
precision muon chambers called MiniDAQ~\cite{MDTelx} was adopted. 
This implementation of the MiniDAQ system was capable of recording 
integrated pulse height and time relative to the trigger for up to 432 
channels with a sub-nanosecond (0.78125 ns) least significant bit.

For position measurement scans, where only basic hit information was 
needed, either a Wiener NIMbox~\cite{NIMbox} module, configured as a 
20 channel scaler, or a CAEN V560 16 channel scaler~\cite{CAENV560} 
was employed to acquire data.
In these measurements, signal processing was done in three stages. 
The panel signals were first discriminated. 
One set of discriminator output logic signals were injected into the scalers.
A second set of discriminator signals were logically OR'ed, and used to 
generate a 1~$\mu$s long veto which was returned to the discriminators
with a measured delay of 60 ns. 
This veto blocked counts in the scaler from the occurrence of later  
pulses that might occur in a panel.

For each of the above dedicated data acquisition methods, analysis software 
was written
using LabVIEW~\cite{LabVIEW} and C++/ROOT~\cite{ROOT}.

\subsection{Radioactive sources}

Because of the low rate of cosmic ray muons through the instrumented area 
of the panels, a number of measurements reported here were made by using 
radioactive $\beta$  emitters: $^{90}$Sr and $^{106}$Ru.
These measurements probed the panel sensitivity to the applied 
voltage by measuring hit rates, and gauged the position resolution.
The $^{90}$Sr source (3.7 mCi), produces, at the end of its decay chain 
a $\beta$ spectrum with a maximum 
energy of 2.28~MeV~\cite{Sr90}. A $^{106}$Ru source, about $3 \; \mu$Ci at the
time of the experiments here reported, yields a $\beta$ energy spectrum end 
point of 3.54~MeV~\cite{RU106}.
The sources were separated from the active pixel volume by various air gaps 
and the 2.23~mm glass plate of the PPS, resulting in substantial beta energy 
loss.
Measurements with radioisotopes were self-triggered by the logical OR of 
all the readout channels after discrimination.

\section{Measurements}

The measurement results reported here included evaluation of the response 
of type VPA and MP panels (see Table~\ref{table:PANEL_SPEC}) filled with 
various gas mixtures using Ar, CO$_2$, CF$_4$ and, for thermal neutron 
detection, $^3$He. 
The panel response to radiation, background hit rates and the occurrence of 
discharge spreading were investigated using  low energy $\beta$ particles, 
$\gamma$ rays, thermalized neutrons from radioactive sources and cosmic ray 
muons.  
Efficiency and 
time response for a few gas mixtures and pressures were 
probed using cosmic ray muons.
Position reconstruction and spatial resolution were explored using a 
slit-collimated radioactive source whose position was set  with a 
computer-controlled  servo-motor arm.

\subsection{Response to $\beta$ sources}
\begin{figure}[!ht]\centering
  \includegraphics[width=0.8\textwidth]{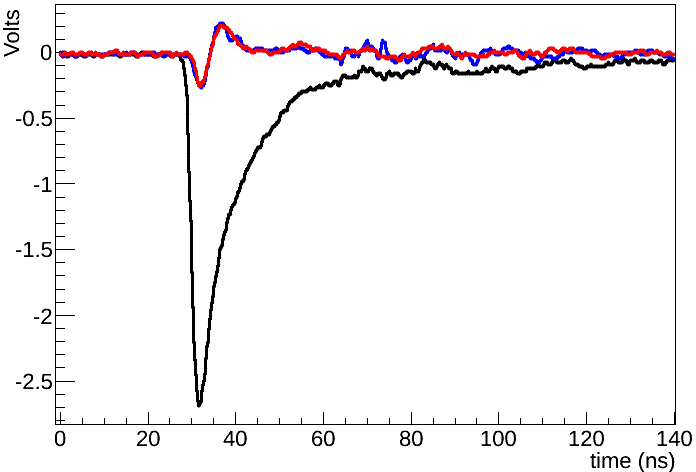}
  \vspace{-3mm}
  \caption{A representative signal induced in a type MP panel 
           (in Table~\ref{table:PANEL_SPEC}), 
           filled with 90\% Ar and 10\%~CF$_4$ at 600~Torr. 
           The large  trace shows the negative discharge pulse.
           Other  traces show transient activity on the 
           nearest neighbor readout electrodes 
           (see Figure~\ref{fig:SPICE_pulse} for comparison). 
           Rise time was about 2 ns. }
  \label{fig:PulseSignal1}
\end{figure}

All panel types responded to the radiation emitted from $^{90}$Sr and 
$^{106}$Ru sources, with all of the tested gases.
Panels were tested at pressures ranging from as low as 200~Torr to 
slightly below atmospheric pressure, because
the tested PDPs were not designed for positive pressure. 
This paper reports measurements made at pressures from 600-730~Torr.
Figure~\ref{fig:PulseSignal1} shows a representative signal induced by 
a $^{90}$Sr source irradiating an MP type panel. 
Similar signals were observed using $\beta$ sources in a panel 
filled with Xe at 600~Torr~\cite{OURPPSPAPER1}, 
sealed seven years before the observation~\cite{PPS_NIM}.

\begin{figure}[!ht]\centering
  \includegraphics[width=0.8\textwidth]{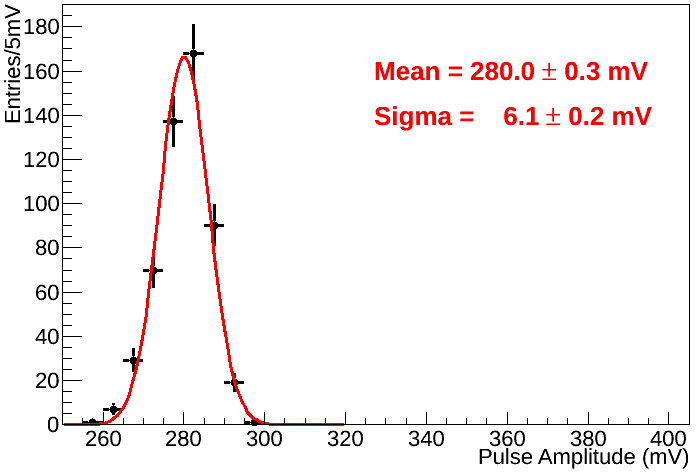}
  \vspace{-3mm}
  \caption{Distribution of the amplitudes, attenuated by 20dB, for a 
           panel  (type VPA in Table~\ref{table:PANEL_SPEC})
           filled with 99\%~Ar and 1\%~CO$_2$ at 600~Torr operated at 860~V.
           The Gaussian fit shows that the dispersion around the 
           mean is 2\%.}
  \label{fig:PulseSignal2}
\end{figure}

The signals from all the tested gases were characterized by large amplitudes 
of 1-10 V volts that depended highly on the gas content 
and panel type, as well as fast rise times, around 1 to 3 ns.
These general features were also evident in the SPICE signal simulation shown 
in Figure~\ref{fig:SPICE_pulse}.
The large signals did not require amplification and sometimes required 
attenuation.
For each test configuration of panel type, gas mixture, pressure and high 
voltage, the induced signal amplitudes were uniform, with approximately 2\% 
variation.  This can be seen in one case in Figure~\ref{fig:PulseSignal2} and 
is expected for a Geiger type discharge. 

\begin{figure}[!ht]\centering
  \includegraphics[width=0.8\textwidth]{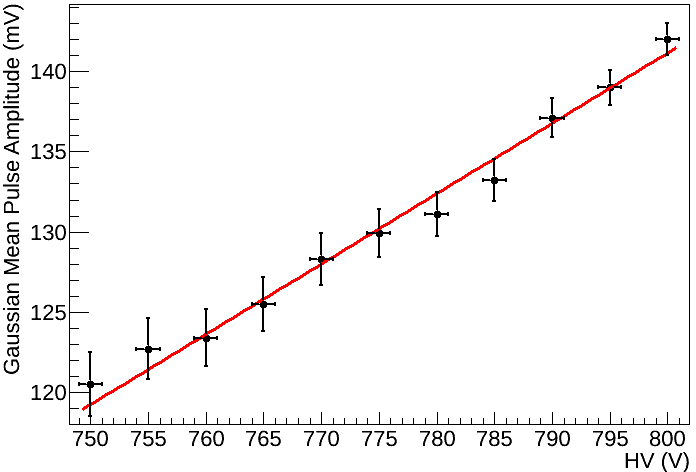}
  \vspace{-3mm}
  \caption{Gaussian mean of the signal amplitude, attenuated by 20 dB, 
           vs HV for a panel (type MP in Table~\ref{table:PANEL_SPEC})
           filled with 10\%~CF$_4$ in Ar at 600~Torr.}
  \label{fig:AmplitudeVsHV}
\end{figure}

Two experiments were conducted to validate the COMSOL and SPICE models.
In the first, a type MP panel (see Table~\ref{table:PANEL_SPEC})  filled 
with a gas mixture of 10\%~CF$_4$ in Ar at 600~Torr was used.  
The dependence of the signal amplitude on the applied voltage was measured, 
the operating voltage range was chosen for maximum signal rate with 
minimum source-free (background) rate. 
This was expected to be linear, based on the idealized capacitive
discharge model expressed by the circuit in Figure~\ref{fig:signal_extraction1}.
The waveform of a single pixel pulse at increasing HV was recorded. 
The signal amplitude  was then fitted with a Gaussian function.
Figure~\ref{fig:AmplitudeVsHV} displays the dependence of the means
on the  applied HV, confirming the expected linear relationship.

\begin{figure}[!ht]\centering
  \includegraphics[width=0.8\textwidth]{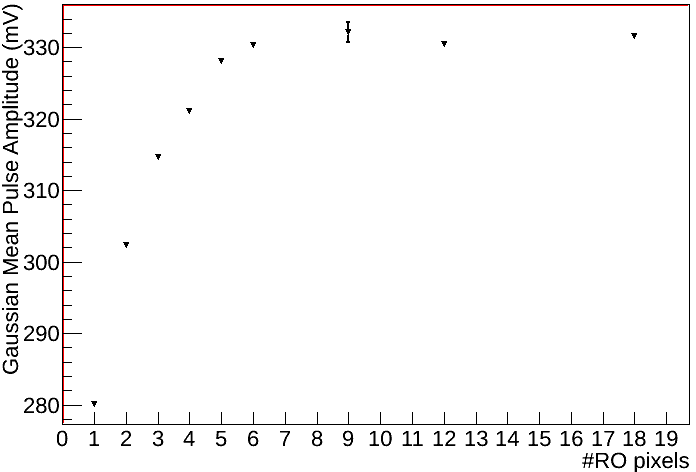}
  \vspace{-3mm}
  \caption{Gaussian mean of the signal amplitude vs. number of connected
           readout lines. VPA panel filled with 10\%~CF$_4$ in Ar at 600~Torr.}
  \label{fig:PulseAvsNro}
\end{figure}

In a second experiment, a type VPA panel (see Table~\ref{table:PANEL_SPEC})
filled with a gas mixture of 10\%~CF$_4$ in Ar at 600~Torr was used at fixed 
operating voltage of 1100V. The signal amplitude dependence on the number 
of readout lines instrumented was measured. 
During these measurements, unconnected electrodes were left floating. 
The exponential  behavior of the data in Figure~\ref{fig:PulseAvsNro} 
shows that with five or more lines the amplitude of the signal reached 
an asymptotic value, consistent with the use of a $5\times5$ matrix for 
the SPICE electrical simulations.

\subsubsection{Voltage scan}
\begin{figure}[!ht]\centering
  \includegraphics[width=0.8\textwidth]{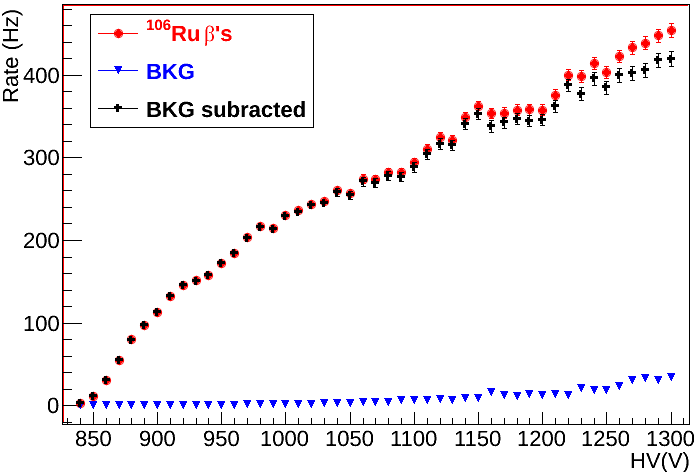}
  \vspace{-3mm}
  \caption{Response to $^{106}$Ru source  (circles),  background rate
           (triangles) and signal with  background subtracted 
           (crosses)  vs. the applied voltage.
           Shown is the combined rate of four pixels in a panel (type VPA 
           in Table~\ref{table:PANEL_SPEC})  filled with 1\%~CF$_4$ in Ar 
           at 600~Torr.}
  \label{fig:VoltageScan}
\end{figure}

The PPS response to radioactive sources and the associated background 
(i.e. no source) were measured over a range of applied HV.
Figure~\ref{fig:VoltageScan} presents an example of a voltage scan taken with 
a gas content of 1\%~CF$_4$ in Ar at 600~Torr. In this measurement one HV 
line was quenched with a 44 M$\Omega$ resistor and four readout channels were 
instrumented.  For every applied voltage two measurements were taken, 
one in which a $^{106}$Ru source was positioned in a fixed location above 
the panel's active area, and one with the source removed.

These measurements provided a relative rate dependence on the HV.
The net rate appeared to reach a plateau at about 1150~V, extending 
for about 60 volts. The dead time is around 350~$\mu$s, so in this 
region the maximum fraction of hits potentially lost because of the 
dead time is below 12\%.
Importantly, the source-free rate remained very low for a large range of 
HV before gradually increasing at higher values of the applied HV.  
The source-free hit rate was primarly attributed to discharges resulting 
from spontaneous formation of ion pairs in the gas, from electron surface 
emission due to stochastic collisions and from photoelectric processes.
At higher applied HV the probability for spontaneous ionization and
regeneration increased leading to a higher background hit rate.

\subsubsection{Quench resistor dependence}

\begin{figure}[!ht]\centering
  \includegraphics[width=0.8\textwidth]{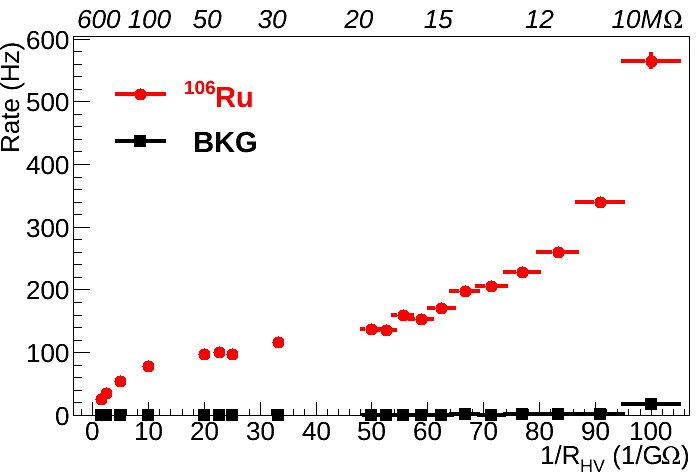}
  \vspace{-3mm}
  \caption{Signal induced by $^{106}$Ru source (circles) and background rate 
           (squares) dependence of the reciprocal of the line quenching 
           resistor. Panel is type VPA filled with 1\%~CO$_2$ in Ar at 
           600~Torr, operated at 815~V.}
  \label{fig:RQuench_dependance}
\end{figure}

For a given panel and gas mixture, a characteristic response curve was 
generated, giving the dependence of the source induced hit rate on the 
HV quench resistance. This is shown in Figure~\ref{fig:RQuench_dependance}, 
plotted as a function of the reciprocal of the line quench resistor. 
For this measurement the panel was filled with 1\%~CO$_2$ in Ar at 600~Torr 
and was operated at 815V. 
The radioactive source was $^{106}$Ru and the hits were collected from four 
readout electrodes crossing a single HV line. 
The quench resistors cover the range from 10-600~M$\Omega$.

This measurement showed that for resistance values below 20~M$\Omega$ the 
recovery time was insufficiently long to prevent the formation of afterpulses. 
These afterpulses increased the measured hit rate.
Conversely, for resistance values above about 100~M$\Omega$ the  
pixel's RC constant was too high for the applied source intensity. 
This was because the time between successive hits from the source was less
than the pixel recovery time, so that the observed hit rate saturated, being 
limited by the RC constant. For this specific gas mixture, pressure and source 
intensity, the working range of quench resistances was obtained where the 
measured hit rate was flat.

\begin{figure}[!ht]\centering
  \includegraphics[width=0.8\textwidth]{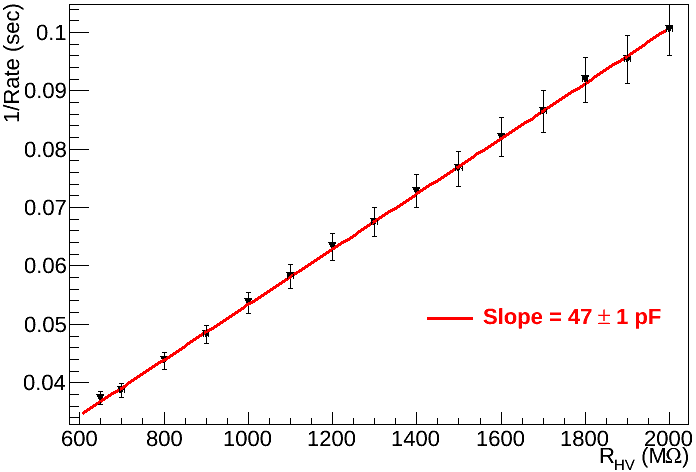}
  \vspace{-3mm}
  \caption{Net hit rate of four instrumented pixels at very high R$_{quench}$.}
  \label{fig:HighR}
\end{figure}

These measurements were extended to very high quench resistor values, 
shown in Figure~\ref{fig:HighR}, using exactly the same setup parameters
as in Figure~\ref{fig:RQuench_dependance}. 
This measurement probed the recovery time constant 
$\tau = {\alpha \cdot RC}\/{N_P}$ where
$\alpha$  is the number of RC time constants needed for a cell's recovery, 
$ R $ is the quench resistance, 
$ C $ is the pixel's effective capacitance (described earlier) and 
$N_P$ is the number of instrumented pixels.  
The slope of the linear fit to the data was $ 47 \pm 1 $~pF. 
Using the measured value for the capacitance, $C = 2.7 \pm 1$~pF and 
$ N_P = 4$, then $\alpha = 4.3 \pm 1.6$. This result suggested that, for the 
cell to become fully active after a discharge, the electric field must have
returned to within a few percent of its pre-discharge strength.
For a 44~M$\Omega$ external quench resistance, as used for the
experiment in Figure~\ref{fig:VoltageScan}, the recovery time for a 
single HV line is around 500~$\mu$s.

\subsection {Measurements with cosmic ray muons}

Cosmic muons were used to explore the PPS response to minimum  ionizing 
particles (MIPs). The main features investigated were the efficiency and 
timing response. Uniformity was also measured. For these measurements an
external plastic scintillator trigger was used.
This trigger provided the necessary coincidence and timing of signals 
induced in the panel to cosmic muons.
\par
\begin{figure}[!ht]\centering
  \includegraphics[width=0.8\textwidth]{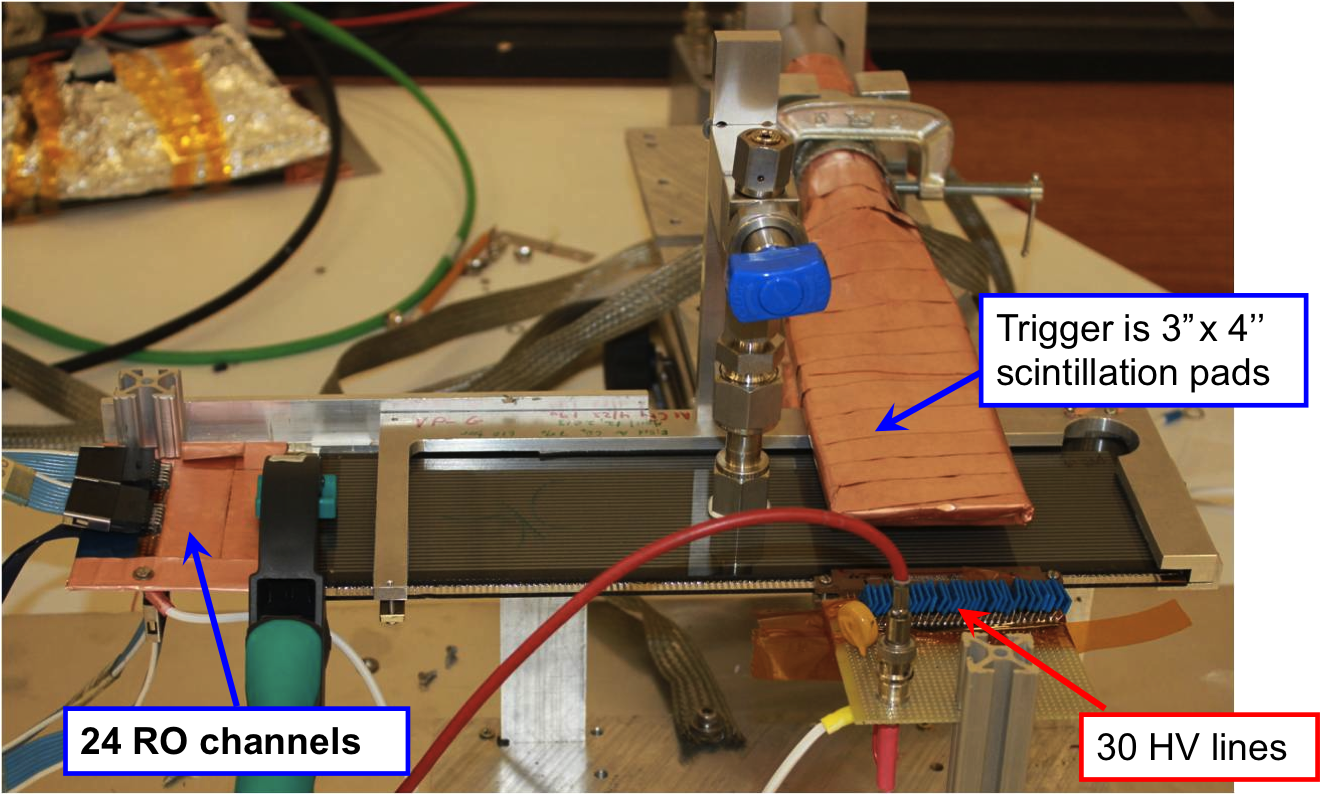}
  \vspace{-3mm}
  \caption{Trigger setup with the panel in a sandwich of scintillators.}
  \label{fig:Set-up}
\end{figure}

The trigger hardware was formed by two scintillator paddles 
(7.5 cm $\times$ 10 cm) placed above and below the 
instrumented areas of the panel (see Figure~\ref{fig:Set-up}).

The trigger signal was the time coincidence of the discriminated pulses 
(20 ns wide, NIM logic signals) from the two scintillator photomultipliers.
The coincidence window of the panel and trigger was 2~$\mu$s. 
The uncorrelated single count hit rate in each scintillator was tens of Hz, 
so the rate of accidental coincidences was negligible.

\subsubsection{Uniformity of response}

Commercial PDPs are fabricated with tolerances not necessarily as stringent 
as required for detectors.
As more HV and readout lines were instrumented, variations in the count 
rates were observed on different parts of the panels. 
A study was performed to obtain a baseline measurement of the uniformity 
of the panel as a function of position.

\begin{figure}[!ht]\centering
  \includegraphics[width=0.8\textwidth]{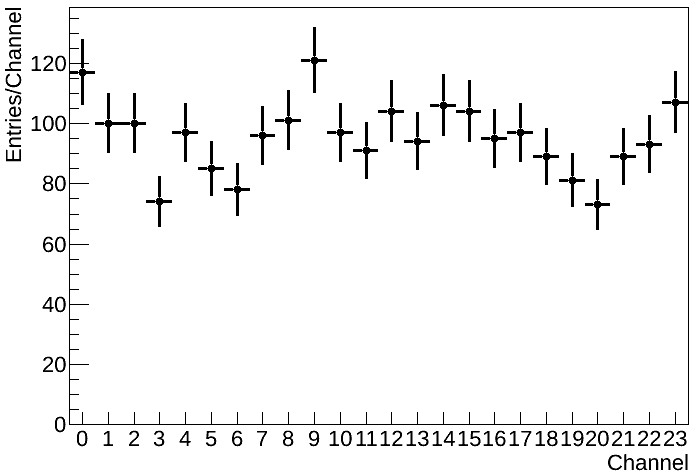}
  \vspace{-3mm}
  \caption{Hit map of cosmic rays acquired with a panel 
           (type VPA in Table~\ref{table:PANEL_SPEC}) filled 
           with 1\%~CF$_4$ in Ar at 730~Torr and operated at 1040V. 
           This panel had 24 readout and 30 HV lines instrumented for 
           this experiment.}
  \label{fig:Uniformity}
\end{figure}

Figure~\ref{fig:Uniformity} shows the distribution of cosmic ray muon
PPS hits per readout line summed over the 30 HV lines instrumented. 
For this panel, the RMS fluctuation around the mean of 95.4 is less than 12\%.
This variation, however, reflects the convolution of the panel non-uniformity 
with the trigger spatial non-uniformity, evaluated to be no more than 10\% 
across the instrumented area of the panel.
These variations are expected based on the non-uniformities of the
thick film electrode printing technique used in PDP construction.

\subsubsection{Secondary pulses} 
\label{DischargeSpreading}
Secondary pulses, referred to here as secondaries, are defined as those 
pulses occurring after the primary one either on the same pixel (afterpulses) 
or in a different location as the result of discharge spreading. 
Discharge spreading is primarily caused by both drifting metastable 
species and by VUV photons propagating to nearby pixels.
The results of the position scans (see Section \ref{PositionScan}) suggested 
that detrimental effects of secondaries are limited. 

\begin{figure}[!th]\centering
  \includegraphics[width=0.8\textwidth]{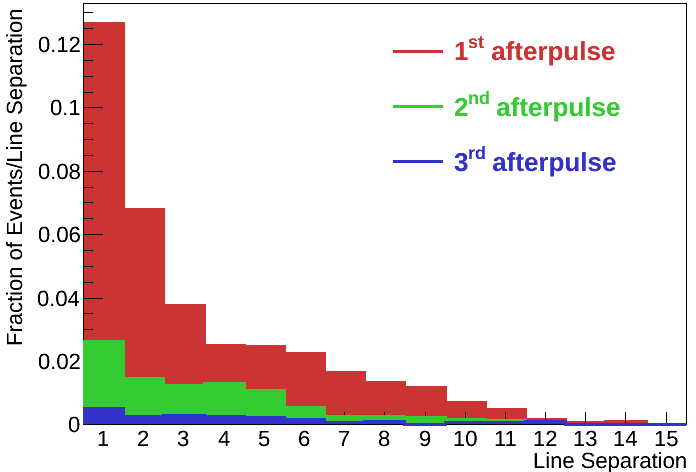}
  \vspace{-3mm}
  \caption{Secondary  pulse readout line separation
           between the first  hit of the event and the n$^{th}$ hit in time.
           The gas is 1\%~CF$_4$ in Ar at 730~Torr in VPA panel.}
  \label{fig:DL}
\end{figure}

\begin{figure}[!th]\centering
  \includegraphics[width=0.8\textwidth]{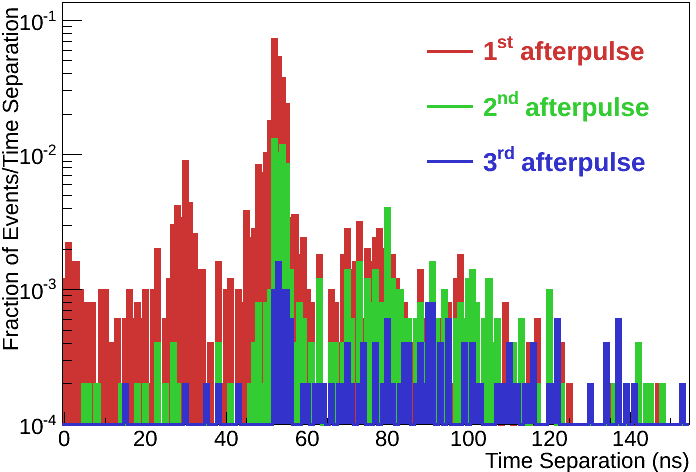}
  \vspace{-3mm}
  \caption{Secondary pulse time separation between the n-th hit 
           of the event and the trigger hit for 1\%~CF$_4$ in Ar at 730~Torr.}
  \label{fig:DT}
\end{figure}

The frequency of secondaries is dependent on the panel geometry, high 
voltage, fill gas and pressure. The frequency and position distribution 
of secondaries was investigated with a VPA panel filled with a gas mixture 
of 1\%~CF$_4$ in Ar at 730~Torr using cosmic ray muon data. 
Figures~\ref{fig:DL}~and~\ref{fig:DT} show the position and time separation 
of secondaries measured using this panel.
The data contained about 5000 muon triggered events of which 64\% have a 
single hit (no afterpulse) and the remainder contain two or more hits.
\par
A similar measurement was done on a type MP panel with 10\% CF$_4$ in Ar 
at 730 Torr at 910 V.  Here the number of events with secondaries on any 
pixel neighboring the triggered readout pixel in a 1~$\mu$sec window  
was directly counted on a digital sampling oscilloscope.
The fraction of events with secondaries was 3\%. 

The number of secondary ($1^{st}$, $2^{nd}$, $3^{rd}$, etc.) hits in 
a line decreased with distance from the primary hit location as shown in 
Figure~\ref{fig:DL}.
This shows that afterpulses and the discharge spreading are decreasing 
rapidly after the third line 7.5~mm away from the primary pulse. 
Although the secondary hits appeared mostly on a time scale of 100 ns or 
less, as shown in Figure~\ref{fig:DT}, this did not preclude longer time 
scale processes such as metastable states or ion drift from initiating 
discharges.

\subsubsection{Arrival time}

The arrival time of the signal relative to the trigger was measured.
Arrival~time=0 corresponds to the passage of the muon through the panel.
The important parameters of the signal time distribution are the mean, 
i.e. the discharge formation time, and the width, i.e. the fluctuation 
in the discharge formation time, that limited the timing resolution of 
the detector.

\begin{figure}[!ht]\centering
  \includegraphics[width=0.8\textwidth]{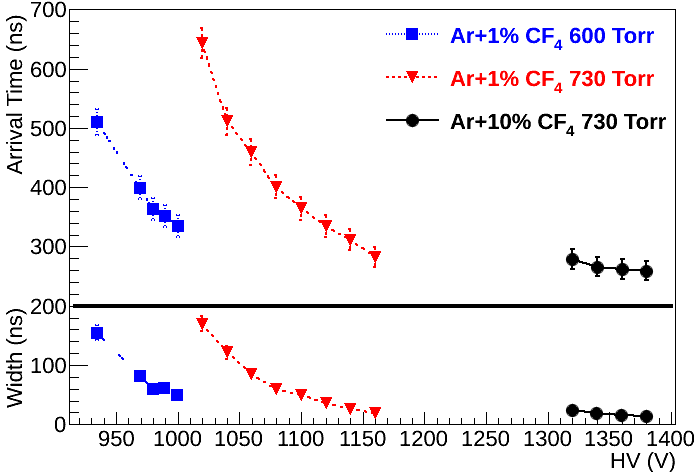}
  \vspace{-3mm}
  \caption{Compilation of cosmic muon mean (top) and Gaussian width (bottom) 
           of the arrival time distribution in various gas conditions. 
           No trigger time subtracted (see text).}
  \label{fig:arrival_time_comp}
\end{figure}

Cosmic ray muon data were acquired with different gas mixtures and pressures, 
and over a range of applied HV. Figure~\ref{fig:arrival_time_comp} shows the 
mean and Gaussian width of the muon arrival time distribution for various 
operating conditions. 

The arrival time resolution, represented by the Gaussian width
decreased as the HV increased.
In these measurements no trigger timing data were available, so the 
effective time resolution was widened by the 25 ns least count timing
of the data acquisition. In any case the effect of voltage with various
gas mixture and pressures is clearly visible in 
Figure~\ref{fig:arrival_time_comp}: the gas mixture with higher 
concentration of CF$_4$ yields faster timing.

\begin{figure}[!ht]\centering
  \includegraphics[width=0.8\textwidth]{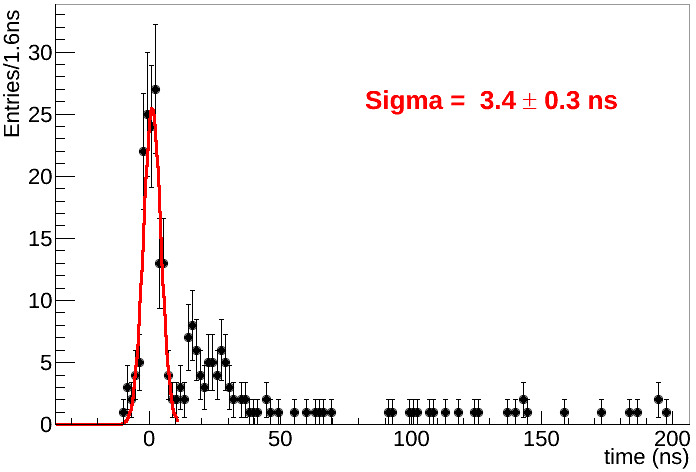}
  \vspace{-3mm}
  \caption{Cosmic ray muons arrival time distribution.
           Gas content: 20\%~CF$_4$ in $^3$He at 730~Torr and 1035~V.
           Trigger cable and electronic delay times accounted for a total
           of 69 ns with an error estimated to be ${} \pm 10 $~ns.}
  \label{fig:CosmicMuon_ArrT}
\end{figure}

Additional measurements were performed in which trigger time information
was available. They displayed a trigger timing jitter of 
$ 1.8 \pm 0.1 $~ns measured as the sigma of a Gaussian fit of the time 
difference between the two trigger scintillators.
Figure~\ref{fig:CosmicMuon_ArrT} shows a cosmic ray muon arrival time 
distribution made with this trigger time subtraction.
This measurement used a gas mixture of 20\%~CF$_4$ in $^3$He at 730~Torr 
operated at 1035 volts.
An intrinsic time resolution of $ 2.9 \pm 0.3$~ns 
was determined by fitting the primary arrival time peak,
and after accounting for the trigger timing jitter.

\subsubsection{Cosmic ray muon efficiency estimation}

A panel's sensitivity to MIPs can be estimated from  cosmic ray muon detection
experiments. For this purpose three definitions of efficiency were used:
\begin{itemize}
\item A {\it raw efficiency}, $\varepsilon_{raw}$, 
      was the ratio of the number of detected cosmic ray muons to the 
      number of triggers: 
      \begin{equation}
        \varepsilon_{raw}= {\rm \frac{N (trigger \ \bullet PPS )}{N_{triggers}}}
        \label{eq:Raw_efficiency}
      \end{equation}   
      where $\rm N(trigger \ \bullet PPS)$ was the number of trigger and 
      PPS coincidences.
\item A panel efficiency, $\varepsilon_{panel}$ was defined by and estimated
      from:
      \begin{equation}
        { \rm \varepsilon_{panel} } = \frac{\rm \varepsilon_{raw}}{A}
        \label{eq:panel_efficiency}
      \end{equation}   
      where $A$ was the fractional acceptance, taken as the ratio of the 
      instrumented panel area relative to the total trigger area.
      $ \sim 0.55$.
\item A pixel efficiency was defined and estimated by,
      \begin{equation}
     {\rm  \varepsilon_{pixel} }= \frac{\varepsilon_{\rm panel}}{f_{pack}}
        \label{eq:pixel_efficiency}
      \end{equation}   
      where $f_{pack}$ was the pixel packing fraction given in 
      Table~\ref{table:PANEL_SPEC}. 
\end{itemize}

\begin{figure}[!th]\centering
  \includegraphics[width=0.8\textwidth]{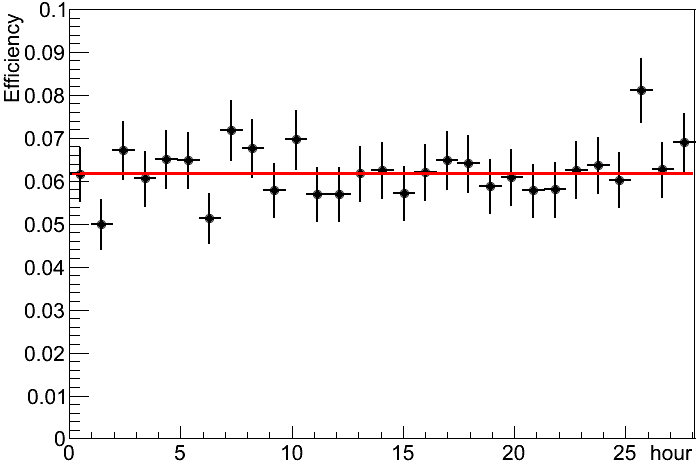}
  \vspace{-3mm}
  \caption{Example of panel efficiency for a panel, 
           type VPA in Table~\ref{table:PANEL_SPEC},  filled with 
           1\%~CF$_4$ in Ar at 600~Torr at an applied voltage of 1000~V.}
  \label{fig:PanelEfficiency}
\end{figure}

The {\it panel efficiency} of these modified display panels was about 
6\% to 13\%, depending upon the HV, limited by the very thin gas gap and 
the small pixel packing fraction. 
Figure~\ref{fig:PanelEfficiency} shows an example of the panel efficiency
for a gas mixture of 1\%~CF$_4$ in Ar at 600~Torr.  
These data showed a stable muon detection rate over a 28 hour run. 

The {\it pixel efficiency} was the efficiency for a single pixel to discharge 
when an ionizing particle traverses it.
A few considerations affected  the estimation of the pixel efficiency:
\begin{enumerate}
\item The trigger rate was uniform across the entire triggering area.
\item  Only the pixel area was active with no significant fringe or 
       edge effects. 
       The effective pixel size did not increase with the applied voltage. 
       The active area of a panel was determined by the packing fraction, 
       estimated for a VPA panel to be 23.5\% (Table~\ref{table:PANEL_SPEC}).
\item The average number of primary ion pairs, N$_P$, was 
      Poisson distributed.  Using Ar as the host gas with 
      10\%~CF$_4$~\cite{PrimariesOnAr} and a gas gap of $385 \; \mu$m, 
      a cosmic ray muon passing through the tested panel produced an 
      average of 0.95 ion pairs.
      The Poisson probability for a MIP to produce at least one ion pair
      in the panel was:
      \begin{equation}
	  P(N_P > 0) = 1- e^{-0.95} \approx 0.61
	  \label{eq:poisson}
      \end{equation}
\end{enumerate}
The maximum possible efficiency for detecting a MIP in a single cell of 
these panels was 0.61, limited by the probability to generate at least 
a single ion-pair in the thin gas gap. 

\begin{figure}[!th]\centering
  \includegraphics[width=0.8\textwidth]{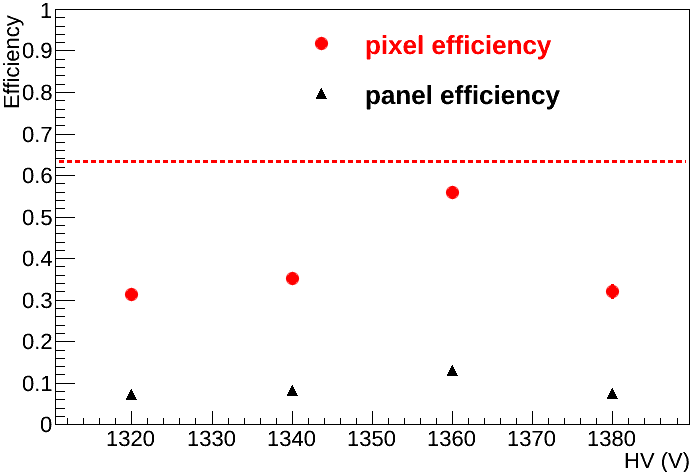}
  \vspace{-3mm}
  \caption{Efficiencies for a type VPA panel, filled with 90\% Ar and 
           10\%~CF$_4$  at 730~Torr. 
           (Triangles) --panel efficiency $\varepsilon_{panel}$ for the 
           instrumented region;
           (Circles)-- single pixel efficiency $\varepsilon_{pixel}$;
           (Dashed line)-- the maximum possible single pixel efficiency for
           this panel's gas pressure and gap size.
           The statistical error bars are smaller than the data points.}
  \label{fig:CorrectedEfficiency}
\end{figure}

The panel efficiency and the single pixel efficiency based on the above 
assumptions are shown in Figure~\ref{fig:CorrectedEfficiency} as a function 
of operating voltage.

For the VPA panel, filled with 10\%~CF$_4$ in Ar at 730~Torr and operating 
at 1360 volts, the pixel efficiency was 55\%, which was nearly 90\% of the 
maximum possible efficiency of 61\%.
The observed efficiency rise with voltage is expected.
The drop off at 1380V is a consequence of dead time caused by numerous pixels 
firing spontaneously at very high rates.

\subsection{Position measurements and spatial resolution}
\label{PositionScan}

In order to investigate the panel's position sensitivity and estimate 
position resolution, a series of position scan measurements were done. 
In these measurements the relative position of a collimated radioactive 
source with respect to the PPS readout electrodes was reconstructed. 
A large working area was instrumented consisting of 20 readout lines 
$\times$ 30 HV lines.
The collimated radiation source produced hit distributions. 
Fits to these hit distributions yielded a measure of the source position.

A computer controlled robotic arm mounted on an X-Y translatable axis 
performed automated position scans. A collimated $^{106}$Ru source was 
attached at one end of the arm and positioned over the panel. 
The collimator consisted of a 2 cm thick graphite block with a 1.25~mm 
slit aperture between the source and glass substrate. 
Control software translated the source in both X and Y directions with a 
precision of about 2~$\mu$m. At each step of the position scan, data were 
acquired for equal time intervals and signals from each readout channel 
were counted with a 20 channel scaler.

\subsubsection{Simulation of the collimated $\beta$ source}
\label{GMC}
\begin{figure}[!htb]\centering
  \includegraphics[width=0.8\textwidth]{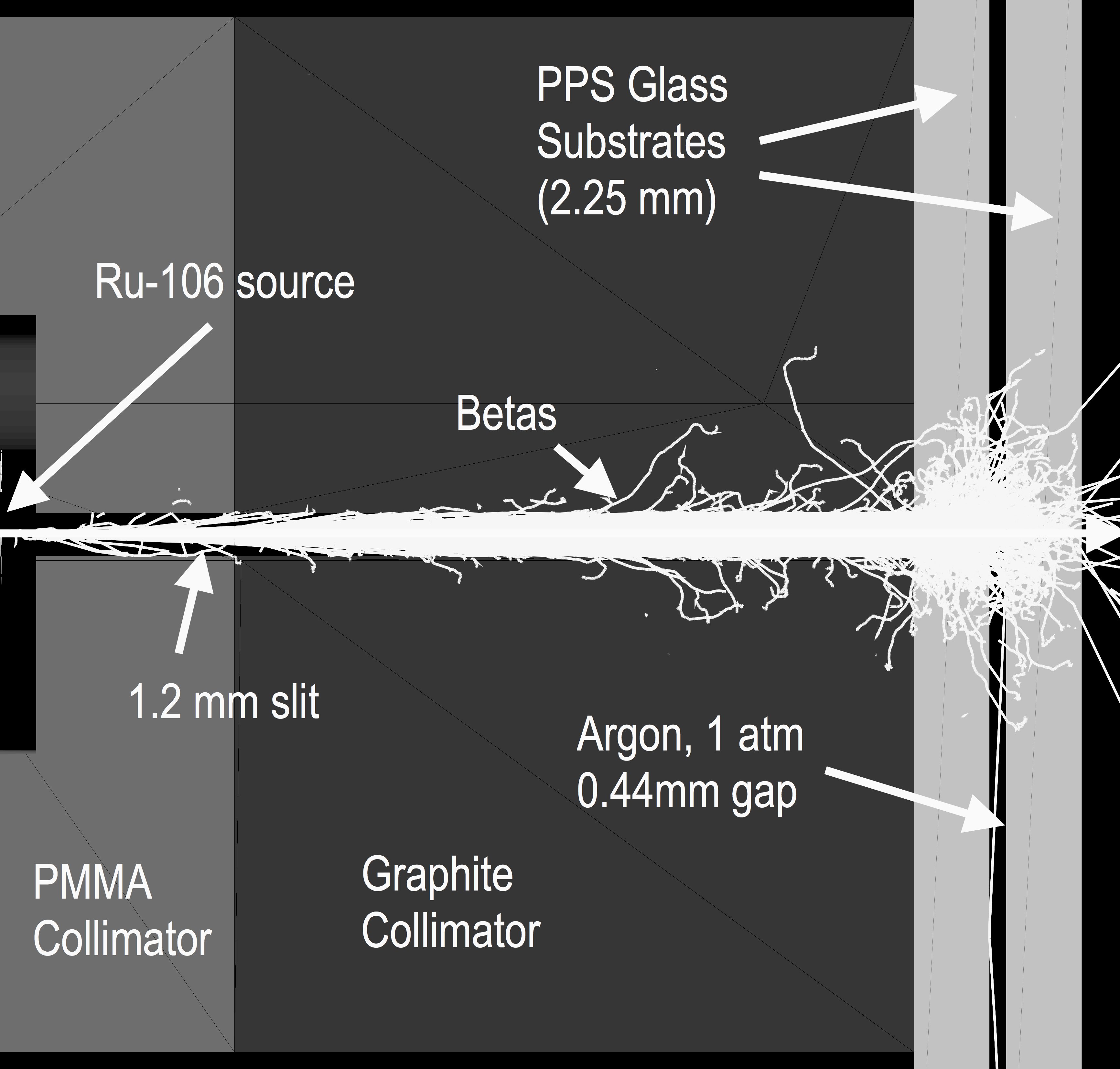}
  \vspace{-3mm}
  \caption{GEANT4 $\beta$ scattering simulation with $^{106}$Ru source.
           For illustration purposes the source is shown as a point 
           source aimed at the PPS.  The beam expands to fill the slit
           because of air scattering, then further diverges in the glass,
           as well as losing intensity. }
  \label{fig:GEANT4_2}
\end{figure}

The investigation of the spatial response of the panel was augmented by a Monte Carlo 
simulation of the $\beta$ source collimation and PPS material scattering.
GEANT4~\cite{GEANT} simulations were used to evaluate the contribution 
of the scattering of the source electrons to the measured distribution of betas in the detector. 
The full simulation started with the $\beta$ energy spectrum of the bare isotope,
included scattering in the ceramic matrix within which the isotope was deposited, the 
collimator material and aperture, and the glass substrates of the panel. The emission of 
$\beta$s was in all directions, sampled appropriately from the energy distribution.
The limited sampling in  Figure~\ref{fig:GEANT4_2} shows the effect of the 
air scattering, the effectiveness of the collimation and the glass energy loss and dispersion, using a
pencil beam sampled from the energy distribution. 
$\beta$s entering the panel 
experienced significant scattering and absorption in the front glass 
substrate, so much so that few were able to exit the panel through the 
back glass substrate.
 
\begin{figure}[!htb]\centering
  \includegraphics[width=0.8\textwidth]{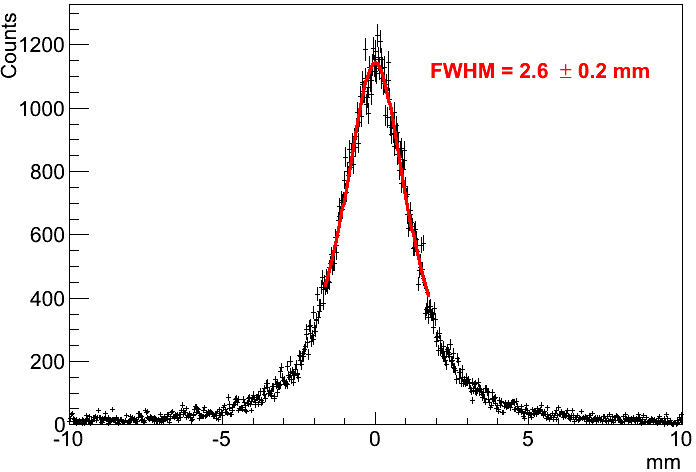}
  \vspace{-3mm}
  \caption{GEANT4 simulation  showing the expected distribution of $\beta$s
           from the slit collimated  $^{106}$Ru source inside the PPS cell
           gas volume. The Breit-Wigner fit was limited to the peak region
           (i.e. the continuous line).}
  \label{fig:GEANT4_1}
\end{figure}

The initial 1.25~mm collimated beam of $\beta$ particles had spread at the 
entrance of the discharge gas volume to a distribution with full width 
at half maximum (FWHM) of about 2.6~mm with long non-Gaussian tails, as 
shown in Figure~\ref{fig:GEANT4_1}. The resulting ``collimated'' $\beta$ 
beam inside the PPS thus also illuminated adjacent sense electrodes on 
each side of the targeted electrode under the graphite slit. 

\subsubsection {Spatial resolution determination from data}

\begin{figure}[!ht]\centering
  \includegraphics[width=0.8\textwidth]{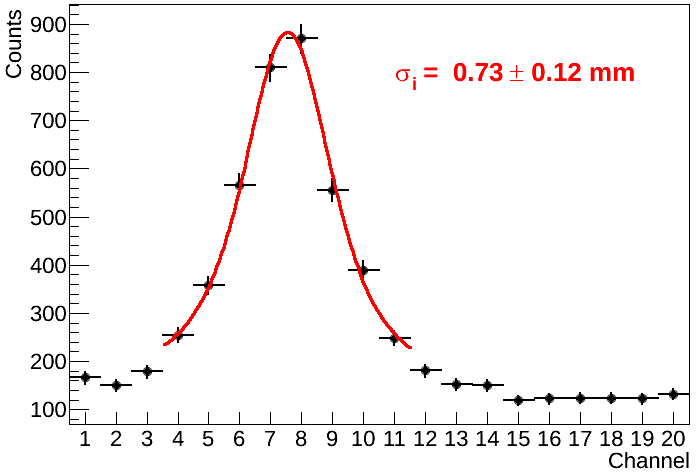}
  \vspace{-3mm}
  \caption{Representative hit distribution induced in a 1~mm pixel pitch 
           panel (type MP in Table~\ref{table:PANEL_SPEC}), filled with
           10\%~CF$_4$ in Ar at 600~Torr by the collimated $^{106}$Ru source.
           The  line represents the convolution fit (see text).
           The Gaussian width is reported.}
  \label{fig:PositionScanHitMapFit1}
\end{figure}

The intrinsic resolution was estimated from the convolution fit of the panel 
hit distribution using a Gaussian to represent the intrinsic resolution and 
a Breit-Wigner function that well described the Monte Carlo distribution around the 
peak region.
Figure~\ref{fig:PositionScanHitMapFit1} displays the result of a single 
position scan run, using the $^{106}$Ru source and acquiring data for 
20 minutes at each point. The intrinsic resolution obtained for the 1~mm 
pixel pitch panel in Figure~\ref{fig:PositionScanHitMapFit1} was
$\sigma_i = 0.73 \pm 0.12$~mm.

This approach yielded an upper limit to the resolution because it ignored 
three systematics.  

One was the absence of secondary $\beta$s in the simulations.
These secondaries originated from the conversion of $\gamma$s 
(e.g., X-rays shown in Figure~\ref{fig:GEANT4_2}). While expected only at 
the few percent level, these $\beta$s produced a very broad distribution 
and were superimposed on the distribution from primary $\beta$s. 

A second systematic effect arose from the lack of timing information 
in the readout method employed for these tests. In this open architecture 
panel about 1/3 of the hits yielded an afterpulse within a few tens 
of $\mu$sec. These afterpulses were mostly vetoed and did not register 
as hits. 
Nevertheless, less than 5\% of the afterpulses incremented a 
scaler channel before the veto was applied.

The third systematic omission was that the simulation generated the 
distribution of $\beta$s at the entrance to the gas volume. This fails to 
account for the large angular distribution of the particles introduced by 
multiple scattering in the air column and in the glass, which would cause 
the distribution to spread significantly as it propagates through the gas 
volume.

\begin{figure}[!ht]\centering
  \includegraphics[width=0.8\textwidth]{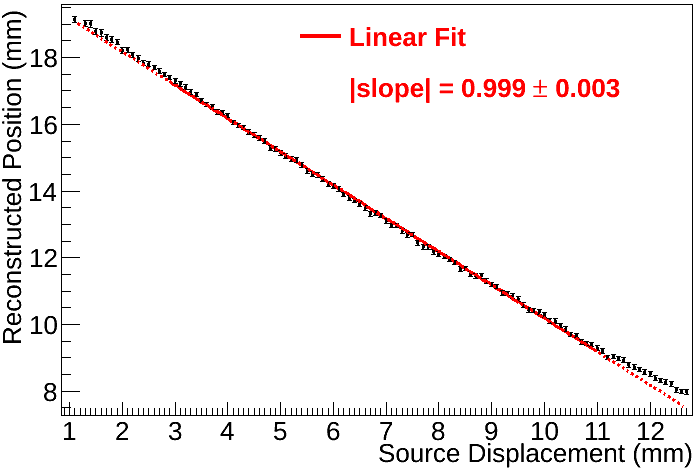}
  \vspace{-3mm}
  \caption{Results of the position scan using a type MP panel 
           (Table~\ref{table:PANEL_SPEC}), filled with 
           10\%~CF$_4$ in Ar. Each data point is the Breit-Wigner fit 
           mean corresponding to the reconstructed position of the source.}
  \label{fig:PositionScanHitMapFit2}
\end{figure}

In general, for each complete scan, the hit map from each step was fitted 
with  a Breit-Wigner plus linear function to model the peak and long tails 
of the distribution.  The position of the source, as seen by the detector, 
was then taken to be the value of the mean of the fit function.
Figure~\ref{fig:PositionScanHitMapFit2} shows reconstructed source positions 
for a series of 100~$\mu$m steps from a single position scan. 

\subsubsection{Medium energy protons}

In addition to the described position resolution results from a $\beta$ 
particle source, similar results were obtained using a proton beam.
A type VPA panel was exposed to a collimated, 226~MeV proton beam 
from an IBA C-235 proton beam therapy accelerator.  An experiment using the 
1~mm diameter beam, with an intensity of $2 \times 10^6$ particles/sec-mm$^2$ 
demonstrated the beam position measurement to be consistent
with the 2.5~mm pixel pitch~\cite{IEEE2012}.

\subsection{Neutron detection}

In collaboration with Reuter-Stokes (General Electric Co.) in Twinsburg, OH, 
a prototype PPS was evaluated as a detector capable of neutron detection with 
a low gamma ray interaction rate. A 2.5~mm pitch
panel was filled with a gas mixture of 80\%~$^3$He and 20\%~CF$_4$ at 730~Torr.
Some 600 instrumented pixels, almost 39~cm$^2$, were exposed to thermal 
neutrons produced by various neutron sources (i.e., $^{252}$Cf, $^{241}$Am-Be 
and $^{239}$Pu-Be).
The sources were encapsulated and nested in the center of a cylinder of high
density polyethylene.

\begin{figure}[!th]\centering
  \centering
  \includegraphics[width=0.8\textwidth]{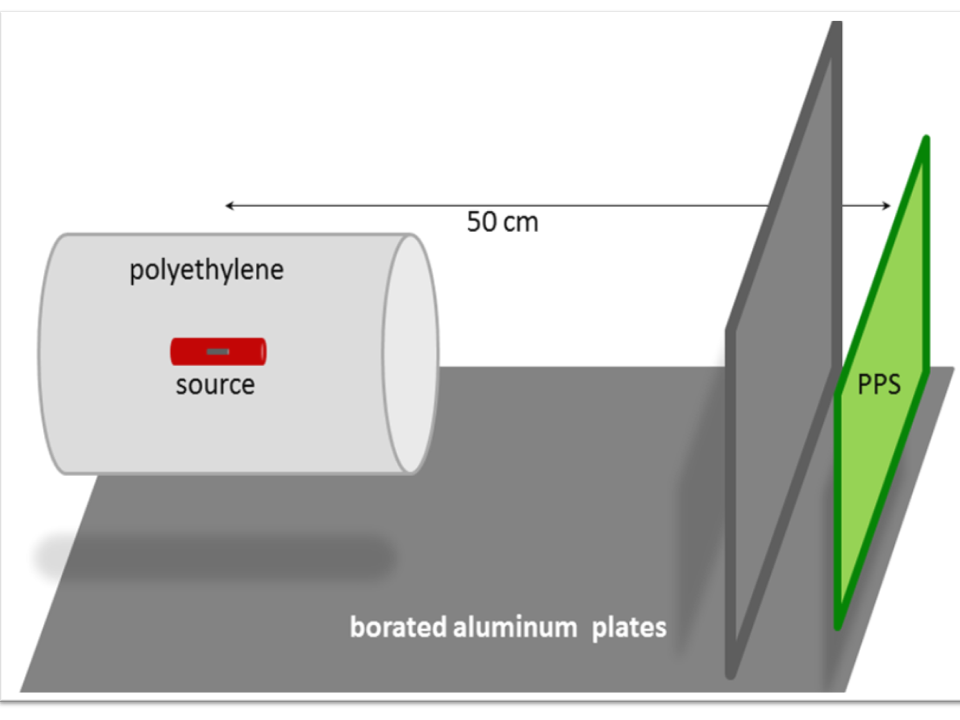}
  \vspace{-3mm}
  \caption{Sketch of setup for neutron measurements.
           The $^{252}$Cf source in a high density polyethylene moderator.
           A $^{10}$B-Al plate is interposed between the source and the 
           panel for some measurements, and removed for others (see text). 
           The underlying wood table surface is also fully covered with
           borated-aluminum to reduce the probability of events 
           generated by neutron scatters.} 
  \label{fig:neutronSetup}
\end{figure}

Hit rate measurements were performed with and without a $^{10}$B-Al
neutron mask between the source and the panel.
The borated aluminum mask was 7.75 mm thick, and had a $^{10}$B areal density 
of $50$ mg/cm$^2$.  $^{10}$B has a cross section of 3835 Barns for thermal 
neutrons. 
A calculation of thermal neutron transmission based on the above values yields
0.001\%.  The transmission of gammas is calculated to be 70\%-80\%.
A GE proprietary measurement determined that at most 0.1\% of the 
source generated neutrons were transmitted by the mask. 

The thermal neutron capture on $^3$He resulted in the emission of a
low energy proton and a triton, sharing a total kinetic 
energy of 764 KeV. Since the interaction occurred essentially at rest, these 
particles were emitted isotropically, with a range of 1 to 2 cm in the gas 
mixture of the panel. They were very highly ionizing, producing many hundreds 
of ion-pairs in the panel gas gap.

\begin{figure}[!th]\centering
  \includegraphics[width=0.8\textwidth]{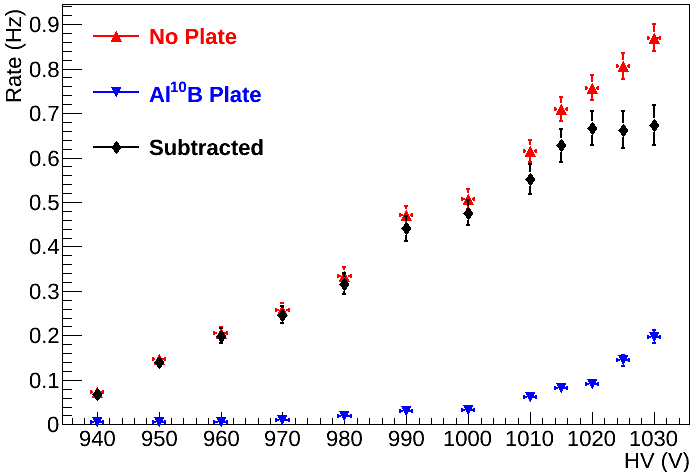}
  \vspace{-3mm}
  \caption{Rates measured by the 20 readout x 30 HV instrumented lines of the
           2.5~mm pitch PPS device (type VPA in table~\ref{table:PANEL_SPEC})
           filled with 80\%~$^3$He and 20\%~CF$_4$ 
           at 730~Torr when exposed to a calibrated $^{252}$Cf source
           at different values of the applied HV.  Triangles-- the total rate; Inverted triangles-- rate with
           a neutron mask between source and panel
           ($\gamma$-component); Diamonds--  the net rate (i.e. neutrons only)
           after subtracting the $\gamma$-component.
           }
  \label{fig:CfVsHV}
\end{figure}
Results of two sets of measurements with and without the neutron source
at different HV values are shown in Figure~\ref{fig:CfVsHV}. 
After subtracting the $\gamma$ contribution to the total rate, a plateau 
was observed in the measured neutron rate above 1015 volts.

The expected thermal neutron hit rate $R_n$ was calculated as:
\begin{equation}
 R_n =\sigma(^3He,n ) \cdot \varepsilon_{det}\cdot  
      \frac {\rho_{He}} {M} \cdot \tau \cdot \Phi_n\cdot A
\label{eq:Rn}
\end{equation}
where $\sigma(^3He,n)$ is the $^3$He thermal neutron absorption 
cross section at 0.025 eV = $5327 \pm 10 $~barns~\cite{neutronxsec};
$A$ was the instrumented panel area 
    ($\rm 20 \times 30 \; pixels \sim 39 \; cm^2$);
$\rm \rho_{He}=1.11\cdot 10^{-4} \; g/cm^3 $ and 
$ \rm M = 5.01\cdot 10^{-24} \; g$ based on the density of ${\rm ^3He}$ 
in the panel under lab conditions; and
$\rm \tau = 385 \; \mu m$ was the thickness of the panel gas gap.
The thermal neutron flux incident on the  instrumented region of panel, 
$\Phi_n$, was determined by a detailed GEANT4 MC calculation, and also 
with a direct measurement. 
The former yielded  a flux of $3.4 \pm 0.01 $ (statistical) Hz~cm$^{-2}$, 
and the latter $3.8 \pm 0.4 $ Hz~cm$^{-2}$. 
The measurement error was dominated by an estimated 10\% systematic
uncertainty.
The measured flux value was used.
The factor $\varepsilon_{det}$  represented the efficiency to detect the  low 
energy breakup fragment passing through the instrumented area of the panel.

Using Eq.~\ref{eq:Rn}  with the above values for the tested configuration 
the calculated rate was $ 0.7 \pm 0.1 \times \varepsilon_{det}$~Hz.
The  measured rate at the last point of Figure~\ref{fig:CfVsHV}  was 
$ 0.67 \pm 0.04 $~Hz.
The comparison of the predicted and measured rates suggests that, within  
an error dominated by the measurement of the thermal neutron flux, the net 
efficiency, $\varepsilon_{det}$,  to detect the captured
neutron  breakup fragments was consistent with unity.  
The absolute efficiency for thermal neutron detection was
\begin{equation}
 \varepsilon_{n} = \frac{R_n}{\Phi_n \cdot A} = 0.005 \pm 0.001 \; (systematic)
\label{eq:effAbs}
\end{equation} 
The error was dominated by systematic uncertainty of the thermal neutron flux 
at the panel. The low efficiency was  due to the minute amount of $^3$He at 
low pressure in the 0.38~mm gas gap. By comparison, a commercial neutron 
counter~\cite{GEneutron} with a 25 mm radius tube and pressurized with 
$^3$He to 4 atm has an  efficiency for thermal neutrons  of  62\%.

Low sensitivity for $\gamma$ detection is a desirable attribute 
of neutron detectors. 
A measurement of the panel efficiency for $\gamma$ 
particles was conducted by directly irradiating the panel with an intense 
$\gamma$ source, $^{137}$Cs with an activity of 8.5 mCi, set at a distance 
such that the panel received a rate of $\approx 3 \times 10^5 $ $\gamma$/sec 
($\pm 10\, \%$) over the instrumented region.
Table~\ref{table:GAMMA_RESULTS} reports the results for the $\gamma$-efficiency 
and corresponding neutron efficiency for a selection of HV settings. 
While this $\gamma$ efficiency is not considered low~\cite{GAOHe3}, 
it can be further reduced  principally by using thinner substrates, 
``thin-film'' $\rm (< 1\,\mu m)$ rather than thick-film 
$\rm (\sim 25 \, \mu m)$ electrodes and by reducing the thickness 
of the dielectric around each pixel.

\begin{table}[!th]\centering
\caption{\footnotesize{\textsl{$\gamma$ irradiation test results. 
The  $\gamma$-efficiency is the ratio of hit rate from the $^{137}$Cs  source 
exposure and the measured $\gamma$ dose rate over the panel area.
The neutron efficiency is the ratio of the number of hits collected at
the corresponding HV and the measured thermal neutron 
rate over the active area of the panel.}}}
\vspace{0mm}
\small
\begin{tabular}{cccc}\hline\hline
       HV(V) & $\gamma$-efficiency & n-efficiency & ratio  \\
	\hline 
        970 & $3.0 \pm 0.8 \cdot 10^{-7}$ & $1.7 \pm 0.3 \cdot 10^{-3}$ & $1.8 \pm 0.8 \cdot 10^{-4}$ \\
       1000 & $3.8 \pm 0.6 \cdot 10^{-6}$ & $3.2 \pm 0.5 \cdot 10^{-3}$ & $1.2 \pm 0.4 \cdot 10^{-3}$ \\
       1030 & $2.5 \pm 0.4 \cdot 10^{-5}$ & $4.6 \pm 0.7 \cdot 10^{-3}$ & $5.5 \pm 1.7 \cdot 10^{-3}$ \\
       \hline\hline
        \end{tabular}
        \label{table:GAMMA_RESULTS}
\end{table}

\begin{figure}[!th]\centering
  \includegraphics[width=0.8\textwidth]{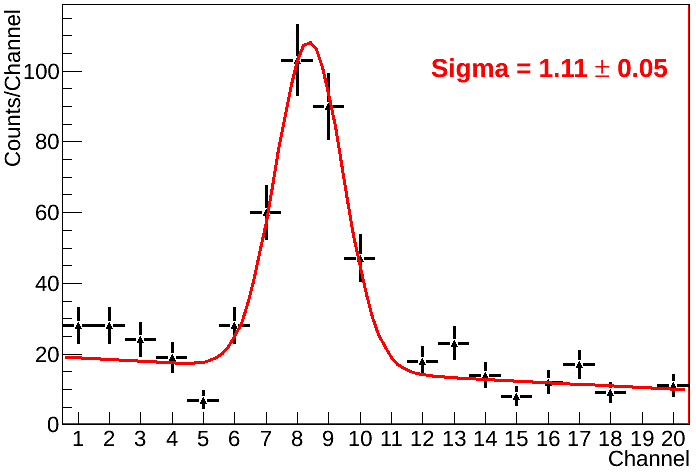}
  \caption{Result of the illumination of a panel 
           (type VPA in Table~\ref{table:PANEL_SPEC}) filled with
           80\%~$^3$He and 20\%~CF$_4$ at 730~Torr at 980~V with an intense
           $^{241}$Am-Be source with a 5~mm collimator.}
  \label{fig:Nslit}
\end{figure}

In a final test intended to evaluate position sensitivity to thermal neutrons, 
a $^{10}$B-Al  neutron mask with a 5~mm wide slit was interposed between the 
neutron source and the panel. The source was located far from the slit 
aperture, at a distance much larger than the slit width, to approximate a 
uniform flux of neutrons and $\gamma$s incident on the panel. 
The resulting distribution in Figure \ref{fig:Nslit} shows a peak over a 
much smaller background.
The Gaussian plus linear fit gives a sigma of 1.11 readout lines, or 2.8~mm. 
The corresponding FWHM is 6.6~mm, in rough agreement with the 5 mm of the 
slit dimension.

\section{Summary}
\label{Summary}

This research was intended to demonstrate the potential of the plasma panel
sensor as an inexpensive, hermetically sealed, scalable, high resolution and 
granularity, fast timing, high performance detector for a variety of sources 
and applications, even in an intense radiation and magnetic field environment.
Test results on prototypes, adapted directly from commercial monochromatic 
plasma display panels, are reported here.
Pulses induced in the panels were systematically characterized, 
expanding on previously reported laboratory 
results~\cite{IEEE2012,SID,IEEE2011,snowmass}. 
The results showed that high gain, fast time response, high spatial 
resolution and high granularity are achievable. 
The first prototype detectors successfully measured minimum ionizing particles,
betas, protons, gammas and thermal neutrons from radioactive sources.
Spatial resolution comparable with the pitch of the device and a time 
resolution as fast as 2 to 3 ns was measured.
We are now investigating devices using discharge cells with better cell 
physical and electrical isolation, a longer drift space and higher fill 
factors. This should lead to a reduced level of discharge spreading, 
lower capacitance and faster discharge times (nanoseconds or lower) 
and very high position resolution. 
Deeper cells with longer interaction paths will increase the detector 
efficiency. Finally, the transition to devices fabricated using much 
thinner substrates will enhance transmission of highly ionizing particles 
while reducing secondary particle and photon generation.





\section*{Acknowledgment}

Development of the PPS project was funded by the U.S. Department of Energy 
(DOE) - Office of Nuclear Physics Small Business Innovation Research  
grant award numbers DE-SC0006204 and DE-FG02-07ER84749 to Integrated Sensors, 
U.S. DOE, Office of Nuclear Physics, Applications of Nuclear Science and 
Technology grant to Oak Ridge National Laboratory, operated by UT-Battelle,
LLC for the U.S. DOE,
and DOE - Office of High Energy Physics grant number DE-FG02-12ER41788 to 
the University of Michigan.  
The research at Tel Aviv University was supported by the I-CORE Program of
the Planning and Budgeting Committee and the Israel Science Foundation 
(grant number 1937\textbackslash 12). 
Funding for scientific exchange and collaboration between Tel Aviv University 
and the University of Michigan was provided by the Israel-American Binational
Science Foundation, grant number 1008123.
\par
Disclaimer:  This report was prepared as an account of work sponsored by an 
agency of the United States Government.  
Neither the United States Government nor any agency thereof, nor any of 
their employees, makes any warranty, express or implied, or assumes any 
legal liability or responsibility for the accuracy, completeness, or 
usefulness of any information, apparatus, product, or process disclosed, 
or represents that its use would not infringe privately owned rights.
Reference herein to any specific commercial product, process, or service 
by trade name, trademark, manufacturer, or otherwise does not necessarily 
constitute or imply its endorsement, recommendation, or favoring by the 
United States Government or any agency thereof.
The views and opinions of authors expressed herein do not necessarily state 
or reflect those of the United States Government or any agency thereof.


\end{document}